 \documentclass[final,5p,times,twocolumn,authoryear]{elsarticle}

\usepackage{amssymb}
\usepackage{lipsum}
\usepackage{longtable}
\usepackage{booktabs}
\usepackage{amsmath}
\usepackage{url}
\usepackage{hyperref}
\usepackage{xcolor}

\journal{High Energy Astrophysics}

\begin{document}

\begin{frontmatter}



\title{Long-term X-ray spectral analysis of Cygnus X-1 using AstroSat}


\author[first]{H Lalthantluanga}
\ead{thantluanga073@gmail.com}

\author[second]{Akash Garg}

\author[second]{Ranjeev Misra}

\author[first]{Vanzarmawii Chhangte}

\author[first]{Lalthakimi Zadeng}

\affiliation[first]{
    organization={Mizoram University},
    addressline={Tanhril},
    city={Aizawl},
    postcode={796004},
    state={Mizoram},
    country={India}
}

\affiliation[second]{
    organization={Inter-University Centre for Astronomy and Astrophysics},
    addressline={Ganeshkhind},
    city={Pune},
    postcode={411007},
    state={Maharashtra},
    country={India}
}

\begin{abstract}
We performed a spectral analysis of long-term observations (2016–2019) of the black hole binary Cygnus X-1 using AstroSat, covering its hard, intermediate, and soft states. During soft states, we found that a typical spectrum fitted with a single Comptonization, a thermal disc, and an iron line yields poor fits and unphysically small disk radii ($R_{\rm in}<1R_{\rm g}$). While adding a relativistic disk component (\texttt{kerrbb}) improved the fit but required high color factors ($f_{\mathrm{col}}\sim3.35{-}5.08$), indicating additional Comptonization. We therefore adopted a dual-Comptonization model with hot (optically thin) and warm (optically thick) coronae. This provided consistent good fits across all soft and hard states of Cygnus X-1. The spectral evolution reveals the critical role of dual Comptonization in shaping the broadband emission of Cygnus X-1.
\end{abstract}




\end{frontmatter}




\section{Introduction}
\label{introduction}

Black hole X-ray binaries (BHXRBs) offer a unique opportunity to study accretion dynamics in extreme environments. In these systems, a companion star loses material that forms an accretion disk around the black hole. As the material in the disk loses angular momentum due to friction, it slowly drifts inward, heating up in the process. This heat energy is predominantly released as X-rays, providing a window into the accretion physics \citep[for example.,][]{frank2002accretion, remillard2006xray}.

BHXRBs are classified as either transient or persistent based on their flux activity throughout their lifetime. Transient BHXRB systems are known for experiencing recurrent outbursts that can last from several months to years before entering a period of quiescence. During these outbursts, they typically transition through different spectral states, categorized as low-hard state (LHS), hard-intermediate state (HIMS), soft-intermediate state (SIMS), and high-soft state (HSS) using the hardness intensity diagram (HID) and spectro-temporal properties \citep[for example,][]{homan2001correlated, homan2005evolution, remillard2006xray, nandi2012accretion, sreehari2018observational, baby2020astrosat, kushwaha2021astrosat}.

Among the various known BHXRBs, Cygnus X-1 (hereafter, Cyg X-1) is one of the most studied and significant sources. Discovered in 1964, Cyg X-1 is the first stellar-mass black hole ever identified and is a persistent, high-mass X-ray binary (HMXB) source \cite[]{bowyer1965cosmic}. The companion star is an O9.7 Iab supermassive star, HDE 2268968, having an inclination angle of 27° \cite[]{orosz2011mass} and an orbital period of 5.6 days \citep[][]{gies1986optical}. The mass of Cyg X-1 was previously estimated to be between 10 and 15 times that of the Sun, but recent measurements suggest it could be as much as 21 solar masses \cite[]{millerjones2021cygnus}. Further, the distance of Cyg X-1 is estimated to be around 2 kpc \cite[]{reid2011parallax}. The dimensionless parameter describing the spin of the black hole has been determined to be exceedingly extreme, at a* $>$ 0.998 \citep[for e.g.,][]{zhao2020spin,kushwaha2021astrosat}.

\begin{figure*}
    \hspace*{-0.07\textwidth}
    \includegraphics[width=1.10\textwidth]{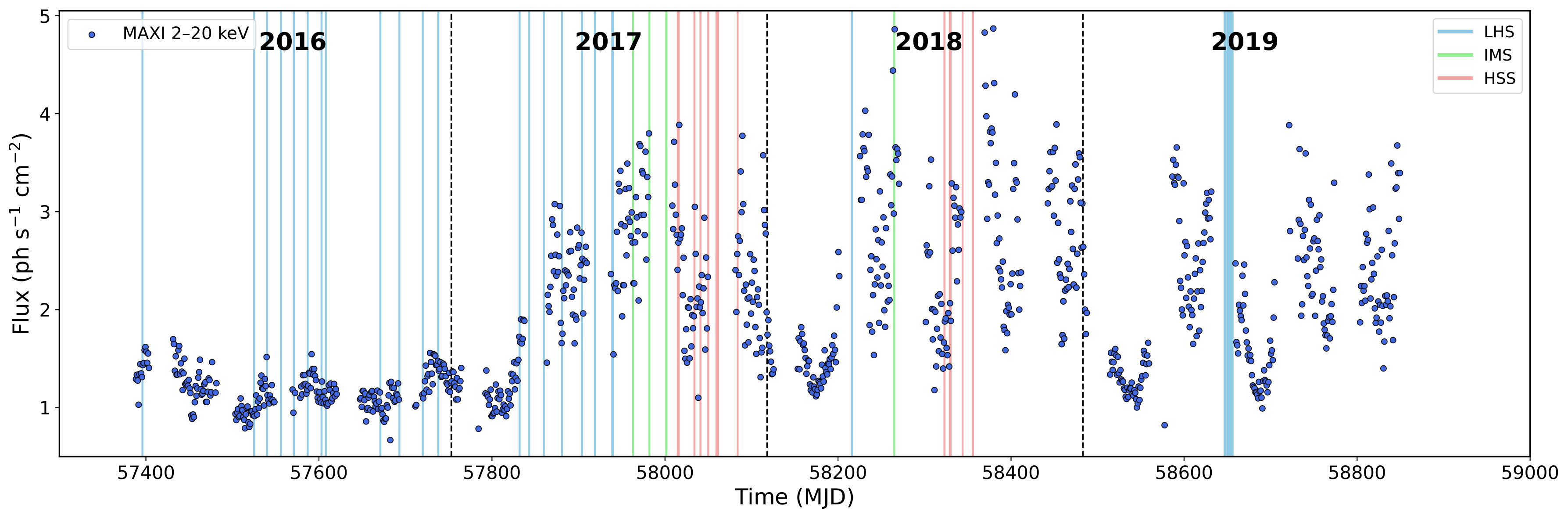}
    \caption{MAXI 2.0–20.0 keV light curve of Cygnus X-1. The vertical colored lines mark the epochs of the AstroSat observations, with blue, green, and red corresponding to the LHS, IMS, and HSS, respectively. The dashed vertical lines indicate the boundaries between the observing years (2016–2019).}
    \label{fig:figure1}
\end{figure*}

Cyg X-1 remains stable in the hard state for a long time, with a $>$ 85\% probability of lasting over 200 hours, while the soft state has a $\sim$75\% chance, whereas intermediate states are short-lived, typically lasting only a few days, requiring monitoring with a resolution of less than one day \citep[]{grinberg2013states}. In the LHS, the X-ray spectrum of Cyg X-1 is characterized by a (cut-off) power-law distribution with a photon index $\Gamma$ $\lesssim$ 2.0 \citep[e.g.][]{tomsick2008broadband}; and the dominant emission is the product of Comptonization events in a hot plasma environment, where the temperature \(kT_e\) is $\approx$ 50-100 keV \citep[for e.g.,][]{haardt1993anisotropic, dove1997corona, zdziarski2003correlations, ibragimov2005broadband, cangemi2019integral, zhou2022spectral}. On the other hand, in the HSS, thermal emission dominates the X-ray spectra from an accretion disk, and the power-law energy spectrum becomes steeper with a photon index $\Gamma$ $\gtrsim$ 2.5 \cite[]{cui1997temporal}. In between these two states, the intermediate state (IMS) is defined, which exhibits variability and spectral properties halfway between those of the hard and soft states \citep[e.g.,][]{zhou2022spectral,belloni1996intermediate}.

The spectral transitions in Cyg X-1 are known to be mainly driven by changes in the X-ray corona's physical properties. In the hard state, emission comes from a hot, optically thin plasma ($kT_e \sim 50\mbox{--}150~\mathrm{keV}$) with an optical depth of $\tau \sim 1$, where soft disc photons are up-scattered via inverse Comptonization \citep[e.g.,][]{zdziarski2020spectral, misra2017astrosat}. The corona is thought to be a compact inner flow or a patchy layer above the truncated accretion disc. As the system transitions to the soft state, the corona cools, becoming less prominent as the disc extends inward, increasing seed photon flux and resulting in a softer spectrum. The coronal temperature decreases to $kT_{\mathrm{e}} \lesssim 40$ keV, while optical depth increases, potentially forming a hybrid thermal-nonthermal electron distribution \citep[e.g.,][]{delsanto2013magnetic, gierlinski1999soft, walton2016soft}. In the soft state, the disc dominates the luminosity, with the corona contributing a steep high-energy tail. Intermediate states involve rapid changes in the corona's temperature, optical depth, and size, occurring on short timescales \citep[e.g.,][]{churazov2001soft, kushwaha2021astrosat}. These coronal variations are crucial to understanding Cyg X-1's evolution across different accretion states.

Besides the direct components like thermal emission and a hard Compton tail, the energy spectrum of Cyg X-1 also shows a reflected component. The reflection, which is caused by hard X-rays penetrating the inner region of the optically thick accretion disk, is mostly composed of a characteristic high-energy continuum peaking at roughly 20–30 keV and a fluorescent Fe line at roughly 6-7 keV, depending on the disk's ionization state \cite[]{george1991reflection}.

Broadband spectral fitting of Cyg X-1 has provided key insights into the interplay between thermal and non-thermal emission. Long-term monitoring with RXTE/ASM and CGRO/BATSE shows that as the accretion rate increases, the disk moves inward, enhancing soft-photon feedback that cools the corona and softens the spectrum. In the soft state, the disk extends to the innermost stable orbit, while high-energy emission likely arises from magnetic flares or a dynamic corona above the disk \citep{zdziarski2003correlations}. Similarly, RXTE observations (1999–2004) demonstrate that spectral variability is governed by changes in the disk–corona system, where an increasing soft-photon flux reduces the coronal heating-to-cooling ratio ($\ell_{\mathrm{h}}/\ell_{\mathrm{s}}$), indicating a transition from a hot, extended corona in the hard state to a cooler, more compact corona as the disk moves inward \citep{wilms2006longterm}.

Over the last nine years, Cyg X-1 has been repeatedly observed using \textit{Astrosat}, India's first multi-wavelength space observatory \cite[]{agrawal2002astrosat}. \cite{misra2017astrosat} reported the results of the first set of AstroSat/LAXPC observations of Cyg X-1, taken in January 2016. They found that the source is spectrally hard with the presence of a thermal Comptonization component with a photon index of $\sim$1.8 and an electron temperature above 60 keV. The study also revealed weak reflection and potential disk emission. Later, \cite{maqbool2019stochastic} performed a spectro-temporal analysis of Cyg X-1's hard state using four LAXPC and SXT observations, all taken in 2016. Their broad-band spectral analysis showed that all observations could be described by a single-temperature thermal Comptonization model with additional disk and reflection components. \cite{bhargava2022shots} performed a joint AstroSat and NICER analysis of two observations from July 4–5, 2017, identifying 49 simultaneous shots. Their broadband spectral fitting (0.1–80 keV) indicates a truncated accretion disk at an inner radius of $\sim 6.7 \pm 0.2 R_\mathrm{g}$, with shot-to-shot spectral variability driven by inward and outward motion of the disk edge at nearly constant accretion rate, producing a soft X-ray peak around $\sim$2 keV. More recently, \cite{kushwaha2021astrosat} reported the spectral and timing analyses of nine LAXPC and SXT observations of Cyg X-1, taken between 2016 and 2018. They identified an exceptionally soft spectral state in 2017, where the disk emission dominated over the Comptonized component, indicating that the accretion disk extended down to the innermost stable orbit around a near-maximally spinning black hole (a* $>$ 0.9981).

In this work, we present the spectral analysis of a broader set of AstroSat observations of Cyg X-1 taken during 2016-2019. We aim to extensively explore spectral evolutions of Cyg X-1. Our paper is organized as follows: In Section 2, we have given details of the observations and discussed the data-reduction methodology of LAXPC and SXT. In Section 3, we discuss the spectral analysis of Cyg X-1. Lastly, we have discussed the results in Section 4.

\begin{figure}
    \hspace*{-0.05\textwidth} 
    \includegraphics[width=0.55\textwidth]{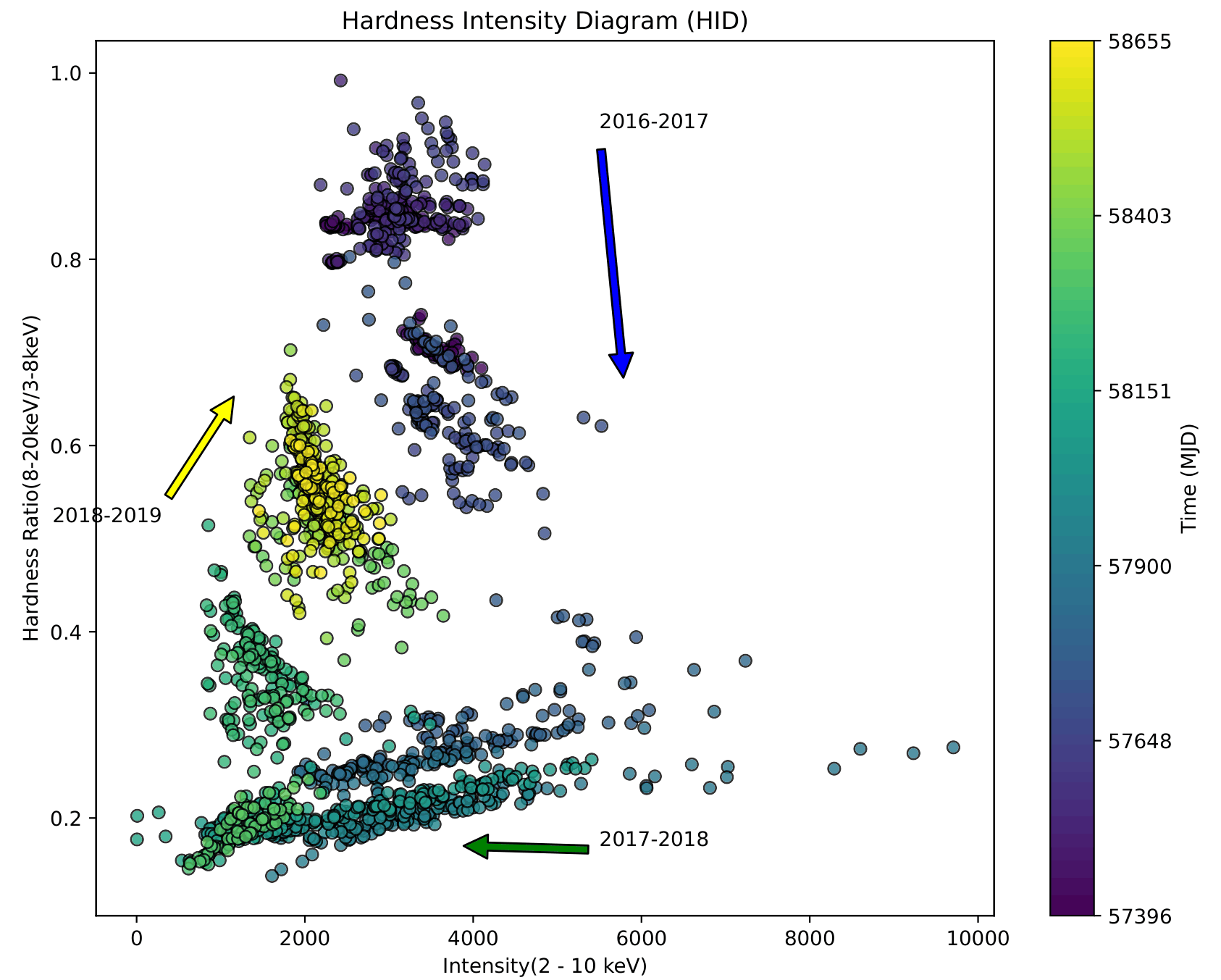}
    \caption{A hardness intensity diagram (HID) for Cyg X-1, where the hardness ratio (the ratio of count rates between 8-20 keV and 3-8 keV) is plotted as a function of the count rate using LAXPC lightcurves from various observations during 2016-2019.}
    \label{fig:figure2}
\end{figure}

\section{Astrosat observations and data reduction}

The continuous monitoring of the MAXI lightcurve in the 2.0-10.0 keV energy band clearly reveals the behavior of this source from 2016-2019 (see Figure \ref{fig:figure1}). Using the hardness ratio of two energy bands, 8.0 - 20.0 keV and 3.0 - 8.0 keV, Figures \ref{fig:figure1} and \ref{fig:figure2} indicate that the source remained in a persistent low-hard state (LHS) throughout 2016 and then began transitioning into a high-soft state (HSS) in 2017. Further, until 2019, it has remained predominantly in the HSS, with frequent transitions to the LHS through the IMS.

During 2016--2019, AstroSat (both LAXPC and SXT) conducted $\sim$47 pointings of Cyg X-1\footnote{\url{https://www.tifr.res.in/~astrosat_laxpc/laxpclog/lc-hdr.html}}. From these, we selected 37 observations for this work, which are marked with vertical black lines in Figure~\ref{fig:figure1}. We further divided these observations into finer segments for spectro-temporal analysis, as shown in Table~\ref{table:tableA1}. Segments~1--12 correspond to the 2016 observations, Segments~13--33 to 2017, Segments~34--41 to 2018, and Segments~42--50 to 2019.

\begin{figure*}
    \centering
    \includegraphics[width=0.998\textwidth]{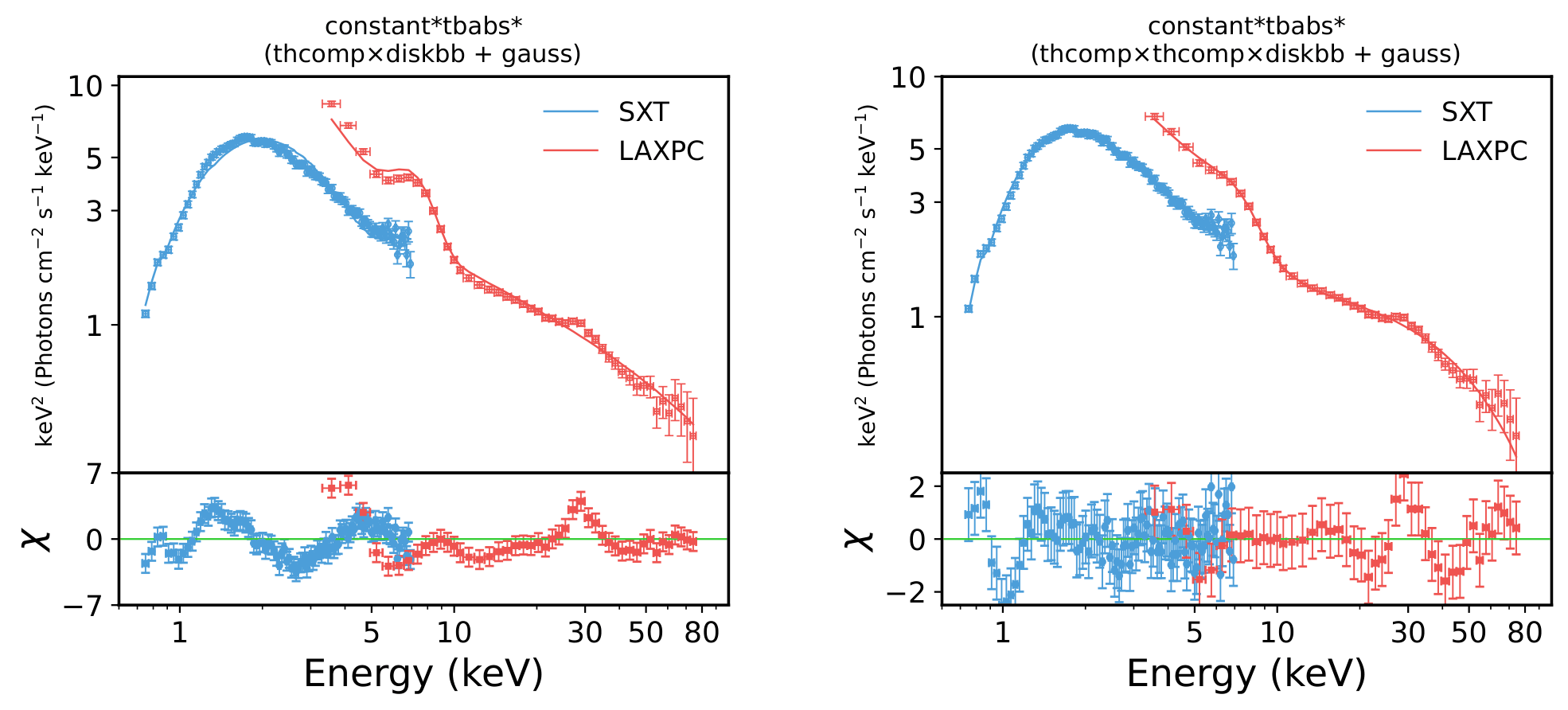}
    \caption{The left and right panels of the upper and lower rows show the spectral fits obtained using single and double \texttt{Thcomp} models, respectively, for the HSS of Segment 28 from the 2017 observations.}
    \label{fig:figure3}
\end{figure*}

\subsection{LAXPC}
The  LAXPC instrument consists of three identical co-aligned X-ray proportional counters (LAXPC10, LAXPC20, and LAXPC30) with a broad spectral range from 3.0-80.0 keV. The counters have a high time resolution of 10 $\mu$s  \citep[e.g.,][]{agrawal2017laxpc, antia2017calibration}. The collective effective area of the three LAXPC detectors amounts to roughly 4500 cm² at 5 keV, 6000 cm² at 10 keV, and approximately 5600 cm² at around  $\sim$ 40 keV \cite[]{roy2019qpo}. We reduced level-1 data to level-2 using LAXPCSOFT software (as of version 15 August 2022) \footnote{\url{http://astrosat-ssc.iucaa.in/laxpcData}}. In addition, we utilized different subroutines within the LAXPCSOFT software package to generate science products such as light curves, energy spectra, and background files. 

\subsection{SXT}

The Soft X-ray Telescope (SXT) aboard the \textit{AstroSat} mission can capture soft X-rays in 0.3-8.0 keV. The effective area of SXT is $\sim$ 90 cm${^2}$ at 1.5 keV \cite[]{singh2014astrosat}. We downloaded level-2 SXT event files from the Astrosat archive and merged all the individual orbits of each observation ID using a Julia script. Then, we used XSELECT to generate the image for the merged event file and chose an annular region to extract the source spectrum. For 2016 observations, we chose an annular region with an inner radius of $1\textquotesingle$  and an outer radius of $12\textquotesingle$ \cite[]{maqbool2019stochastic}. For 2017 observations, we chose an inner radius of $3\textquotesingle$ and an outer radius of $12\textquotesingle$  \cite[]{kushwaha2021astrosat}. The same annular region, with an inner radius of $3\textquotesingle$ and an outer radius of $12\textquotesingle$, was used for both 2018 and 2019 observations. This was done to consider the pile-up effect that the high source flux causes in the charged-coupled device (CCD) of the SXT. The SXT instrument team provided the necessary tools and files for data extraction and analysis, including background, spectral response, and effective area files \footnote{\url{ https://www.tifr.res.in/ astrosat_sxt/dataanalysis.html}}. We used the sxtARFmodule \footnote{\url{ https://www.tifr.res.in/ astrosat_sxt/sxtpipeline.html}}, to correct for any pointing errors in the data and obtain the corrected ancillary response function (ARF). Further, we rebinned the source spectrum using the ftgrouppha command. The final SXT spectrum was created by combining the SXT source binned spectrum with the background, response file (RMF), and corrected ARF files using the grppha command.  

For further analysis, we followed the state classification scheme of \cite{grinberg2013states} based on MAXI light curves, and classified all the segments into LHS, IMS, and HSS. The adopted state boundaries in terms of photon indices obtained from the single Comptonization model (Table~\ref{tab:table1}) differ slightly from those of \cite{grinberg2013states} due to the different spectral models employed.

\begin{table*}[ht]
\centering
\caption{Spectral state classification adopted in the present work using the single Comptonization model.}
\label{tab:table1}
\renewcommand{\arraystretch}{1.25}
\begin{tabular}{lccc}
\toprule
\textbf{Property} & \textbf{Low-Hard State (LHS)} & \textbf{Intermediate State (IMS)} & \textbf{High-Soft State (HSS)} \\
\midrule

Photon index &
$\Gamma < 1.95$ &
$1.95 \leq \Gamma < 2.20$ &
$\Gamma \geq 2.20$ \\

Segments &
1--20, 34, 42--50 &
21--24, 35 &
25--33, 36--41 \\

\bottomrule
\end{tabular}
\end{table*}

\section{Spectral Analysis}
We performed joint broadband spectral analysis of all simultaneous LAXPC 20 and SXT observations. All spectra were fitted using the HEASOFT tool \textit{XSPEC 12.12.1}. While fitting the spectra, a constant factor was added to the model to account for cross-calibration between the SXT and LAXPC. A gain correction was also applied to the SXT and LAXPC spectrum, with the slope fixed at 1, while allowing the offset to vary across all observations. For all the observations, we used the 0.7–80.0 keV energy range.


\begin{figure*}
    \centering
    \includegraphics[width=0.98\textwidth]{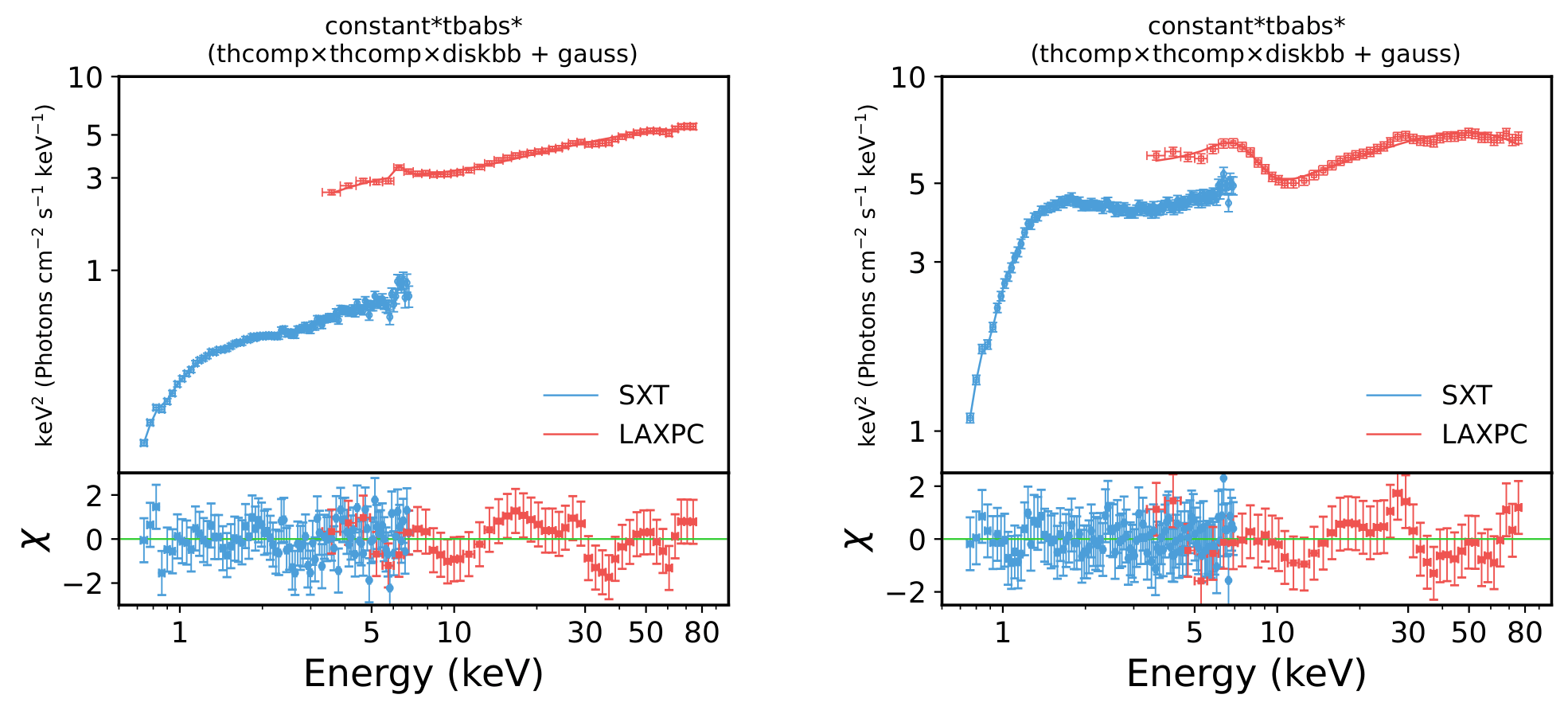}
    \caption{Representative broadband AstroSat/SXT and LAXPC spectrum of Cygnus X-1 obtained during the LHS. The left and right of the upper and lower panel show a comparison fitting for a double \texttt{Thcomp} of Segment 1 and Segment 43 for 2016 and 2019 observations.}
    \label{fig:figure4}
\end{figure*}

Firstly, we began with a HSS spectrum of Segment 28 (i.e, from 2017 observations) and fitted it with a model consisting of interstellar absorption \cite[\textit{Tbabs};][]{wilms2000absorption}, disk emission \cite[\textit{diskbb};][]{mitsuda1984energy}, and a thermal Comptonized component \cite[\textit{ThComp};][]{zdziarski2020spectral}. The spectral fitting model combination of \textit{$\mathrm{constant} * \mathrm{Tbabs} * (\mathrm{ThComp}  * \mathrm{diskbb})$} gave a poor $\chi^2$ value of 987.46 for 101 dof and a systematic error of 3\%. After this, a Gaussian line is taken to account for the Fe~K$\alpha$ emission line. We refitted the data with a model combination of \textit{$\mathrm{constant} * \mathrm{Tbabs} * (\mathrm{ThComp}  * \mathrm{diskbb} + \mathrm{gaussian})$} and obtained a $\chi^2$ value of 495.37 for 139 dof. Notably, we also obtained a disk normalization of $\sim$6486. To evaluate the physical significance of this value, we derived the inner disk radius ($R_{\rm in}$) from the standard definition of the \texttt{diskbb} normalization, given by

\begin{equation}
R_{\rm in,app} = \sqrt{\frac{N}{\cos i}} \left(\frac{D}{10~\mathrm{kpc}}\right)
\end{equation}

where $R_{\rm in,app}$ is the apparent inner disk radius, $D$ is the source distance, and $i$ is the inclination angle.

Following \cite{kubota1998bh}, the true inner radius is given by
\begin{equation}
R_{\rm in} = \xi\, f_{\rm col}^2\, \sqrt{\frac{N}{\cos i}} \left(\frac{D}{10~\mathrm{kpc}}\right)\ \mathrm{km}
\end{equation}

Here, we adopt an inclination of $i = 27^{\circ}$ (corresponding to $\cos i \approx 0.891$) \citep[]{orosz2011mass}, a distance of $D = 2.2~\mathrm{kpc}$ \citep[]{millerjones2021cygnus}, and a correction factor of $\xi f_{\rm col}^2 \approx 1.19$, where $f_{\rm col} \approx 1.7$ and $\xi \approx 0.41$ \citep[e.g.,][]{kubota1998bh, shimura1995hardening}. We obtained physical inner radius of $R_{in}$=22.73 km.

The corresponding gravitational radius is given by
\begin{equation}
R_{g} = \frac{GM}{c^2},
\end{equation}

where $G$ is the gravitational constant and $c$ is the speed of light.  
Substituting $M = 21.2\,M_\odot$, we obtained the gravitational radius of $R_{g}$=31.30 km. This value corresponds to $R_{\rm in} \simeq 0.73^{+0.11}_{-0.11}R_{\rm g}$. Although the inferred radius is small under the standard \texttt{diskbb} assumptions, its exact value is subject to systematic uncertainties associated with the color-correction factor, boundary conditions, inclination, distance, and disk geometry. Therefore, we do not use the inferred radius alone to assess the physical validity of the model. Instead, the inadequacy of the single-Comptonization description is indicated by its poor goodness of fit and by the unusually large color-correction factors required by the \texttt{kerrbb} fits.


In order to understand this small $R_{in}$, we further took a more physical model, relativistic disk emission \cite[\textit{kerrbb};][]{li2005disk}  instead of 'diskbb', and fitted the spectrum with a model combination of \textit{$\mathrm{constant} * \mathrm{Tbabs} * (\mathrm{ThComp}  * \mathrm{kerrbb} + \mathrm{gaussian})$}, where kerrbb norm was fixed at 1.0. We observed a $\chi^2$ value of 389.84 for 139 degrees of freedom and a systematic error of 3\%. During the fitting, the distance, inclination, and mass parameters of \texttt{Kerrbb} were fixed to the values reported for Cyg X-1 \citep{millerjones2021cygnus}. For a spin of \(a_* = 0.98\), the color factor obtained was \(f_{\mathrm{col}} = 3.35\), while for a spin of \(a_* = 0\), we obtained a color factor \(f_{\mathrm{col}}\) to 5.08, with a $\chi^2$ value of 411.80 for 139 dof. The corresponding mass accretion rates were \(\dot{M} = 0.0912 \times 10^{18}\) and \(0.239 \times 10^{18}\ {\rm g\,s^{-1}}\) for the high and zero-spin cases, respectively. Note that we have also explored higher spin values \(a_* = 0.99981, a_* = 1.0\), but the color correction factor remain significantly higher than the typical range of \(1.7{-}2.0\).

The increase in \(f_{\mathrm{col}}\) with decreasing spin arises because a lower-spin black hole has a larger ISCO, producing a cooler disk that requires a higher hardening factor and accretion rate to reproduce the observed flux. The resulting \(f_{\mathrm{col}}\) values (\(\sim3.35{-}5.08\)) are significantly higher than the typical \(1.7{-}2.0\) \citep[]{shimura1995hardening} expected for standard thin disks, suggesting that the thermal component alone cannot explain the soft X-ray emission. This likely indicates the presence of additional emission processes, such as warm Comptonization or disk-corona coupling, contributing to the soft X-ray excess.

Before adding any additional component, we checked for possible deviations from solar abundances in the stellar wind. We replaced the standard absorption model (\texttt{tbabs}) with \texttt{tbfeo}, allowing the oxygen and iron abundances to vary freely. The model combination \textit{$\mathrm{constant} * \mathrm{Tbfeo} * (\mathrm{ThComp}  * \mathrm{diskbb} + \mathrm{gaussian})$} gave a poor $\chi^2 / \nu = 472.35 / 136$. The best-fit values of the abundances came out to be $A_{\rm O} = 0.677$ and $A_{\rm Fe} = 8.38 \times 10^{-5}$ (relative to solar). We also explored the presence of an additional \texttt{ThComp} component. Here, the model combination \textit{$\mathrm{constant} * \mathrm{Tbabs} * (\mathrm{ThComp}  * \mathrm{ThComp} * \mathrm{diskbb} + \mathrm{gaussian})$} gave a good fit with a $\chi^2$ value of 109.80 for 135 dof. We found that one of the two \texttt{Thcomp} components (hereafter, $\mathrm{ThComp}_{l}$) is warmer and has a lower electron temperature than the other \texttt{Thcomp} (hereafter, $\mathrm{ThComp}_{h}$). This indicates a warm corona is required to describe the softer part of the spectrum for Cyg X-1. 

To further understand the physical nature of low KTe component, we estimated its effective optical depth using the known relation between the photon index, electron temperature, and optical depth. In several observations the photon index is pegged at the lower boundary allowed by the model; rather than relying on the best-fit values alone, we therefore estimated the allowed range of optical depth using the confidence intervals of both the photon index and electron temperature, to conservatively account for this boundary effect. We find that the inferred optical depth remains large ($\tau \gtrsim 7$) over the entire allowed parameter space, indicating that the low-temperature Comptonizing component is optically thick. Therefore, while our analysis cannot uniquely distinguish between a warm corona and Comptonization occurring within the upper layers of the accretion disk on geometric grounds, it does show that the observed soft excess is inconsistent with a simple multicolor blackbody (diskbb) alone, and instead requires an additional, optically thick, warm Comptonizing region associated with the inner accretion flow. A comparison of one and two \texttt{Thcomp} models for Segment 28 is shown in Figure \ref{fig:figure3}.

To verify the results for all segments of 2017, we then fitted the remaining spectra (Segments 13 to 33) using the same model combination, finding that the reduced $\chi^2$ values for the best fits range between 0.4 and 1.1. We also found that the spectral index ($\Gamma_h$) varies between 1.62 to 2.248, suggesting that the source was moving through different spectral regimes during this period, likely evolving from the LHS through IMS and into the HSS.

For the HSS of 2016 (Segments 1–12) and the LHS/IMS/HSS of 2018 (Segments 34–41), we applied the same approach as for 2017. The 2016 spectra yielded reduced $\chi^2$ values between 0.4 and 0.9, with $\Gamma_h$ ranging from 1.4–1.7, confirming a typical hard state. The 2018 spectra showed reduced $\chi^2$ values between 0.6 and 1.02, with $\Gamma_h$ between 1.7 and 2.2, corresponding to LHS, IMS and HSS.

All spectra from 2019 (Segments 42–50) yielded reduced $\chi^2$ values between 0.4 and 0.6, with $\Gamma_h$ ranging from 1.70 to 1.79, indicating a return to the LHS. The scattering fraction of the hot corona shows differences between the two epochs, indicating that the 2019 LHS observations are not fully identical to the 2016 ones. A representative LAXPC model comparison of the hard states from 2016 and 2019 is shown in Figure \ref{fig:figure4}.

\begin{figure*}
    \centering
    \includegraphics[width=\textwidth, height= 22.5 cm]{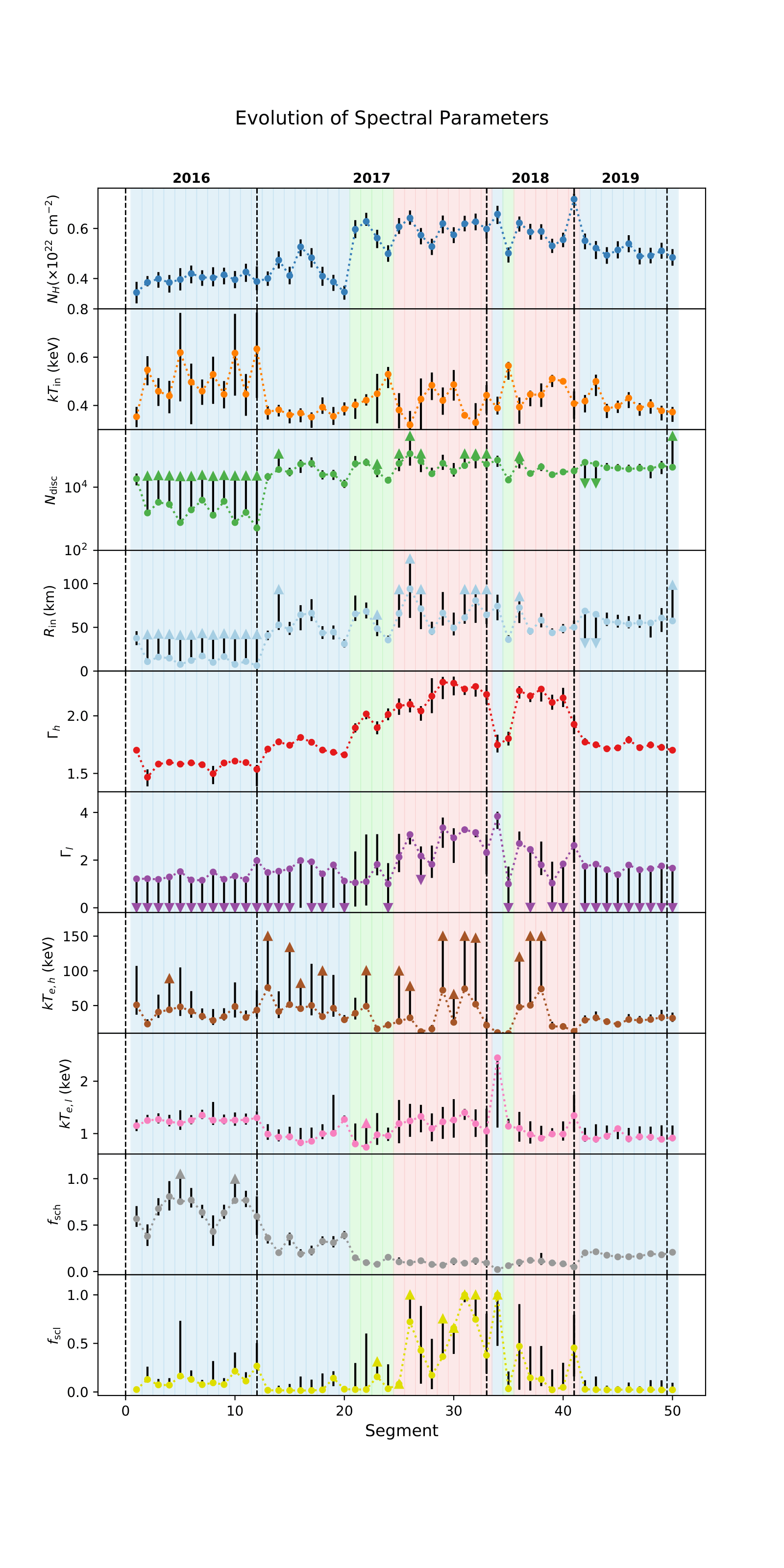} 
    \caption{The time evolution of spectral parameters. The shaded regions indicate the adopted spectral state classification, with light blue, light green, and light red representing the LHS, IMS, and HSS, respectively. The dashed vertical lines separate the observations obtained during 2016, 2017, 2018, and 2019. Points marked with upward or downward arrows denote parameters for which only 90\% upper or lower limits could be constrained.}
    \label{fig:figure5}
\end{figure*}

Table \ref{table:tableA2} lists the best-fit spectral parameters for the fitting. The variation of $\Gamma_h$, $\Gamma_l$, \(kT_{in}\), $N_{disk}$, $N_H$, \(kTe_h\), \(kTe_l\), \(fsc_h\), \(fsc_l\) as well as $R_{in}$ as a function of Segment number (i.e., time) is shown in different panels of Figure \ref{fig:figure5}. 

\section{Discussion and Conclusion}

We conducted long-term spectral studies of Cyg X-1 using Astrosat observations from 2016 to 2019, when the source was in Low-Hard state (LHS), Intermediate state (IMS), and High-Soft state (HSS). We discuss the results below.

The LAXPC HID and the MAXI light curve reveal an evolution between spectral states over the four-year time interval, as shown in Figure \ref{fig:figure1} and \ref{fig:figure2}. Starting from 2016 observations, we found that the hardness lies in $\approx$ 0.8–1.0 with Intensity clustering in $\sim$ 2000 - 4000, showing the typical hard state. As the source goes forward to 2017 observations, the HID shows a clear trend towards softening, with a reduction in the hardness (HR $\approx$ 0.5–0.8), and the Intensity clustering around $\sim$ 3000-5000. By 2018 observations, the source becomes softer with hardness $\approx$ 0.1–0.4 and Intensity in the range $\sim$ 500-9000. Finally, as it moves towards 2019 observations, the hardness increases (HR $\approx$ 0.4–0.9) with Intensity $\sim$ 1000-3500. It is visible through HID and light curve trends that the LAXPC during 2016-2019 has captured the source in LHS, IMS, and HSS, providing an opportunity to study spectral features in these different states of Cyg X-1.

In this work, we began our analysis by examining the soft state spectrum from 2017 observations, by fitting the data with a combination of \texttt{diskbb} and \texttt{ThComp} models. However, this model combination gave a relatively poor fit and an unusually low disk normalization, for instance in particularly for Segment 28, where the derived inner disk radius (\(R_{\mathrm{in}} \sim 22.73~\mathrm{km}\)) corresponded to only about \(0.73\,R_{\mathrm{g}}\), which implies a minimal and unphysical inner disk radius ($R_{in} < 1 R_g$). This implies that the standard \texttt{diskbb} model is insufficient to account for the observed soft X-ray emission.

To further investigate, we substituted the \texttt{diskbb} component with the relativistic disk emission \texttt{kerrbb} model, and this model combination slightly improved the fitting, giving a color factor of ($f_{col} \sim 3.3$) for Segment 28, which is considerably higher than the expected \(1.7{-}2.0\) for a standard thin disk \citep[]{shimura1995hardening}. We extend this analysis to all the soft states' spectra, and we found that $f_{col}$ varies between $\sim 2.8 - 3.5$. These elevated $f_{col}$ values suggest strong disk hardening or additional Comptonization effects. A single Comptonization component alone could not describe the broadband spectrum, so we introduce a second component representing a warm, optically thick region alongside the hot corona. This dual Comptonization model provided a significantly improved fit across all spectral states, suggesting that the broadband emission of Cyg X-1 can be well described by an absorption component, a thermal disk, two thermal Comptonization components, and a Gaussian line fixed at 6.4 keV.

The broadband \textit{INTEGRAL} study of \cite{cangemi2019integral}, which analyzed Cygnus X-1 over the 20--2000 keV energy range, found that the spectrum below $\sim$400 keV is well described by thermal Comptonization, while an additional power-law component is required to account for the high-energy tail above $\sim$400 keV. They also detected polarization of this high-energy component in the hard state, suggesting a non-thermal origin such as jet emission or a hybrid corona. In our analysis, the hot-corona electron temperature ranges from $\sim$10--75 keV, consistent with their thermal Comptonization results. However, the high-energy power-law tail lies beyond the energy range of AstroSat and therefore cannot be constrained in the present work.\\
Recent \textit{AstroSat}/CZTI observations of Cygnus X-1 by \cite{chattopadhyay2024polarization} also reported evidence for an additional high-energy spectral component above $\sim$100 keV. Their spectral modeling of the 30--190 keV data using the \texttt{compps} model showed that the hard X-ray continuum is adequately described by thermal Comptonization, while an additional power-law component is required during the intermediate hard state, consistent with the observed hard X-ray polarization. These results support the broadband picture inferred from \textit{INTEGRAL} study of \citep[]{cangemi2019integral} and suggest that an additional high-energy component may emerge beyond the energy range accessible to our AstroSat LAXPC analysis.

Several studies have shown that a single Comptonization component is insufficient to describe the spectra of black hole binaries, motivating the use of dual Comptonization models. For instance,\cite{basak2017nustar} found, using simultaneous NuSTAR and Suzaku observations of the hard state of Cygnus X-1, found that the spectrum requires an additional soft (warm) thermal Comptonization component in addition to the primary hot Comptonizing corona to account for the soft X-ray excess. More recently, \cite{zdziarski2024black}, using simultaneous NICER, NuSTAR, and INTEGRAL observations of the soft state, likewise required a warm, optically thick Comptonizing layer with kTe of about 1 keV, is in good agreement with the low-temperature component we find in the soft state. \\
However, other multi-mission analyses have modeled the same broadband spectrum using different geometrical assumptions without invoking a second thermal Comptonization component explicitly. \citep[]{walton2016soft}, fitting NuSTAR soft-state spectra, attributed spectral variability primarily to changes in a single hot corona's temperature combined with relativistic reflection from a stable inner disk, rather than a distinct warm layer. Similarly, the Insight-HXMT spectral-timing study of \cite{zhou2022spectral} and the accretion-mode analysis of \cite{feng2022evolution} interpret state-dependent spectral changes chiefly in terms of an evolving corona geometry, moving from a disk-hugging "sandwich" configuration in the hard state to a vertically extended, jet-like geometry in the soft state, with reflection strength as the primary diagnostic, rather than a second Comptonizing zone. This is not necessarily in conflict with our results, since a geometrically evolving single corona and an additional warm layer are not mutually exclusive descriptions, but it does mean that the specific two-thermal-component decomposition we adopt is a modeling choice among several viable ones in the literature, rather than a uniquely established result. This diversity of preferred models across missions likely reflects real degeneracies in broadband spectral fitting — different combinations of corona geometry, reflection strength, and one- versus two-component Comptonization can produce statistically comparable fits to overlapping but non-identical energy bandpasses and instrumental responses.\\
Our AstroSat LAXPC+SXT broadband energy coverage (0.7–80 keV) allows us to constrain the broadband continuum over a wide energy range, and our spectra are best described by the inclusion of an additional warm Comptonizing component to account for the soft excess. This interpretation is supported by similar warm, optically thick Comptonization components reported independently using different missions and different epochs of Cyg X-1 \citep[e.g.,][]{basak2017nustar, zdziarski2024black}, as well as in other black hole binaries such as GX 339–4 \citep[]{dhaka2026gx339}.\\
At the same time, we note that alternative reflection-based and coronal geometry models applied to NuSTAR and Insight-HXMT observations \citep[for example][]{feng2022evolution, walton2016soft, zhou2022spectral} provide comparably good descriptions of their respective datasets without explicitly invoking a second Comptonizing component. This underscores that the spectral decomposition of the broadband continuum of Cyg X-1 remains model-dependent, and that distinguishing between these competing physical scenarios will likely require simultaneous multi-mission observations together with independent constraints from X-ray polarimetry or reverberation timing.

{From Figure \ref{fig:figure5}, the long-term evolution of the spectral parameters between 2016 and 2019 reveals changes in the accretion behavior of Cyg X-1. In 2016, the source remained in a canonical low/hard state (LHS). The absorption column density, ($N_{\rm H}$), varies between  $0.34$-$0.71 \times 10^{22}\,\mathrm{cm}^{-2}$.  The inner disc temperature (\(kT_{in}\)) stayed nearly constant at $\sim 0.4$ keV, consistent with a cool and stable accretion disc. Both photon indices ($\Gamma_{h}$ and $\Gamma_{l}$) were low, as expected in the LHS, and the coronal parameters exhibited minimal variability throughout this period. As the system transitioned from the LHS to the IMS and HSS in 2017, variations in all spectral parameters became evident. The disc temperature increased sharply, indicating enhanced thermal disc emission. During 2018, the source was observed in the LHS, IMS, and HSS and displayed transitions between the IMS and HSS, accompanied by correlated changes in the spectral parameters. By 2019, once the source returned to the LHS, the parameters again stabilized, reflecting a recovery of the LHS configuration. During the 2016 observations, only lower limits on the disc normalization could be constrained. These limits suggest that the normalization may be comparable to, or even larger than, the values measured during 2017–2019. As commonly observed in black hole X-ray binaries, the disc is expected to extend closer to the black hole during softer states, consistent with the higher disc temperatures and spectral softening seen during the transition phases.

\section*{Acknowledgements}
The authors thank the anonymous reviewer for constructive comments to improve the manuscript. We extend our appreciation to the members of the LAXPC instrument team for their invaluable contributions to the development of the LAXPC instrument. We acknowledge the LAXPC Point of Contact (POC) at TIFR for verifying and releasing the data. Analysis in this paper was conducted using LaxpcSoft software. Additionally, we utilized data from the Soft X-ray Telescope (SXT) developed at TIFR, Mumbai, and express our gratitude to the SXT POC at TIFR for verifying and releasing the data through the ISSDC data archive, as well as for providing essential software tools. L.Zadeng acknowledges financial support from the Science and Engineering Research Board (SERB), Department of Science and Technology (DST), New Delhi, Govt. of India vide Lett. No. EEQ/2021/000346.

\appendix

\section{Observation Log and Spectral Parameters}

Table~\ref{table:tableA1} presents the observation log of Cyg~X-1, including the observation time (MJD) and exposure time (ks). Table~\ref{table:tableA2} presents a lists the spectral fitting parameters.
\begin{table*}
\setlength{\tabcolsep}{8pt}
\renewcommand{\thetable}{A\arabic{table}}
\centering
\caption{Summary of the AstroSat observations of Cyg X--1 grouped by year. The spectral state assigned to each observation is based on the single Comptonization model adopted in this work.}
\renewcommand{\arraystretch}{1.0}

\begin{tabular}{c c c c c c}
\hline \hline
Year & Segment No. & State & Obs ID & Time (MJD) & Exposure Time (ks) \\
\hline

     & 1  & LHS  & 9000000258 & 57396 & 24 \\
     & 2  & LHS  & 9000000456 & 57525 & 30.5 \\
     & 3  & LHS  & 9000000476 & 57540 & 29.9 \\
     & 4  & LHS  & 9000000500 & 57556 & 9.13 \\
     & 5  & LHS  & 9000000528 & 57571 & 13.3 \\
2016 & 6  & LHS  & 9000000542 & 57587 & 12 \\
     & 7  & LHS  & 9000000572 & 57603 & 21.4 \\
     & 8  & LHS  & 9000000584 & 57608 & 10.4 \\
     & 9  & LHS  & 9000000722 & 57671 & 16 \\
     & 10 & LHS  & 9000000768 & 57693 & 15.4 \\
     & 11 & LHS  & 9000000834 & 57720 & 17 \\
     & 12 & LHS  & 9000000890 & 57738 & 14.4 \\

\hline \hline

     & 13 & LHS  & 9000001094 & 57832 & 17.1 \\
     & 14 & LHS  & 9000001122 & 57843 & 15.2 \\
     & 15 & LHS  & 9000001180 & 57860 & 10.2 \\
     & 16 & LHS  & 9000001210 & 57881 & 9.11 \\
     & 17 & LHS  & 9000001258 & 57904 & 10.7 \\
     & 18 & LHS  & 9000001304 & 57919 & 9.22 \\
     & 19 & LHS  & 9000001358 & 57939 & 8.15 \\
     & 20 & LHS  & 9000001360 & 57940 & 30.5 \\
     & 21 & IMS & 9000001414 & 57963 & 9.23 \\
2017 & 22 & IMS & 9000001470 & 57982 & 10 \\
     & 23 & IMS & 9000001516 & 58001.530 & 4.35 \\
     & 24 & IMS & 9000001516 & 58001.596 & 4.92 \\
     & 25 & HSS  & 9000001540 & 58014.806 & 54.9 \\
     & 26 & HSS  & 9000001540 & 58015.969 & 42.2 \\
     & 27 & HSS  & 9000001592 & 58034 & 9.59 \\
     & 28 & HSS  & 9000001616 & 58041 & 20.9 \\
     & 29 & HSS  & 9000001636 & 58050 & 21.3 \\
     & 30 & HSS  & 9000001660 & 58059.445 & 33.81 \\
     & 31 & HSS  & 9000001660 & 58060.387 & 32.85 \\
     & 32 & HSS  & 9000001660 & 58061.404 & 39.92 \\
     & 33 & HSS  & 9000001726 & 58084 & 18.9 \\

\hline \hline

     & 34 & LHS  & 9000002024 & 58216 & 20.4 \\
     & 35 & IMS & 9000002120 & 58265 & 20.1 \\
     & 36 & HSS  & 9000002252 & 58323 & 30.2 \\
     & 37 & HSS  & 9000002260 & 58329.312 & 35.55 \\
2018 & 38 & HSS  & 9000002260 & 58329.312 & 37.93 \\
     & 39 & HSS  & 9000002260 & 58330.152 & 33.41 \\
     & 40 & HSS  & 9000002302 & 58344 & 33.7 \\
     & 41 & HSS  & 9000002326 & 58356 & 11.4 \\

\hline \hline

     & 42 & LHS  & 9000002986 & 58646.935 & 23.50 \\
     & 43 & LHS  & 9000002986 & 58647.589 & 20.74 \\
     & 44 & LHS  & 9000002292 & 58649.983 & 25.43 \\
     & 45 & LHS  & 9000002292 & 58650.532 & 38.97 \\
2019 & 46 & LHS  & 9000002292 & 58651.508 & 25.48 \\
     & 47 & LHS  & 9000002292 & 58652.307 & 30.83 \\
     & 48 & LHS  & 9000002292 & 58653.336 & 35.51 \\
     & 49 & LHS  & 9000002292 & 58654.548 & 37.04 \\
     & 50 & LHS  & 9000002292 & 58656.091 & 48.87 \\

\hline \hline
\end{tabular}
\label{table:tableA1}
\end{table*}

\begin{table*}[ht!]
\begin{minipage}{\textwidth}
\centering
\setlength{\tabcolsep}{5pt} 
\renewcommand{\thetable}{A\arabic{table}}
\caption{Best-fitting spectral parameters of Cyg X-1. $N_{\rm H}$: hydrogen column density; 
$\Gamma_{h,l}$ and $kT_{e,h,l}$: photon indices and electron temperatures of hard/soft Comptonization; $fsc_{h,l}$: covering fractions; $T_{\rm in}$, $N_{\rm disk}$: disk blackbody parameters. Errors at $90\%$ confidence; $f$ = frozen parameter.}
\label{tab:Table_SP}
\begin{tabular}{cccccccccccccccc}

\toprule  
\textbf{Sgmt} & \multicolumn{2}{c}{\textbf{Tbabs}} &  \multicolumn{2}{c}{\textbf{Thcomp\_h}} & \multicolumn{3}{c}{\textbf{Thcomp\_l}} &  \multicolumn{2}{c}{\textbf{Diskbb}} & \textbf{$\chi^{2}_{\text{red}}$} \\  

\textbf{No.} & \hspace{0.5cm} $N_H$ ($10^{22}$ cm$^{-2}$) & $\Gamma_h$ & $kT_{e,h}$ (keV) & $fsc_h$ & $\Gamma_l$ & $kT_{e,l}$ (keV) & $fsc_l$ & $T_{\rm in}$ (keV) & $N_{\rm disk}$ ($\times 10^4$) &  \\  

\midrule  
1 & $0.34^{+0.03}_{-0.04}$ & $1.70^{+0.01}_{-0.02}$ & $51.04^{+55.36}_{-14.22}$ & $0.57^{+0.14}_{-0.08}$ & $< 1.21$ & $1.15^{+0.12}_{-0.10}$ & $0.03^{+0.02}_{-0.01}$ & $0.36^{+0.04}_{-0.04}$ & $1.77^{+1.15}_{-0.69}$ & 0.6 \\

2 & $0.37^{+0.02}_{-0.03}$ & $1.46^{+0.06}_{-0.07}$ & $23.38^{+6.71}_{-3.81}$ & $0.38^{+0.12}_{-0.10}$ & $< 1.22$ & $1.25^{+0.11}_{-0.05}$ & $0.13^{+0.13}_{-0.01}$ & $0.55^{+0.05}_{-0.06}$ & $>0.15 $ & 0.6 \\

3 & $0.39^{+0.02}_{-0.03}$ & $1.58^{+0.02}_{-0.01}$ & $40.88^{+25.57}_{-8.72}$ & $0.68^{+0.11}_{-0.07}$ & $< 1.19$ & $1.26^{+0.12}_{-0.07}$ & $0.07^{+0.06}_{-0.01}$ & $0.46^{+0.06}_{-0.05}$ & $> 033$ & 0.6 \\

4 & $0.37^{+0.03}_{-0.04}$ & $1.59^{+0.01}_{-0.01}$ & $> 44.03$ & $0.81^{+0.16}_{-0.14}$ & $< 1.15$ & $1.22^{+0.13}_{-0.08}$ & $0.07^{+0.07}_{-0.01}$ & $0.45^{+0.06}_{-0.07}$ & $>0.28$ & 0.5 \\

5 & $0.38^{+0.04}_{-0.05}$ & $1.58^{+0.02}_{-0.01}$ & $47.99^{+55.81}_{-13.26}$ & $> 0.75$ & $< 1.54$ & $1.18^{+0.28}_{-0.12}$ & $0.17^{+0.56}_{-0.03}$ & $0.62^{+0.19}_{-0.23}$ & $> 0.07$ & 0.9 \\

6 & $0.41^{+0.03}_{-0.04}$ & $1.59^{+0.02}_{-0.03}$ & $ 41.78^{+29.64}_{-9.32}$ & $0.77^{+0.13}_{-0.08}$ & $< 1.17$ & $1.24^{+0.11}_{-0.07}$ & $0.13^{+0.09}_{-0.02}$ & $0.49^{+0.08}_{-0.17}$ & $> 0.19$ & 0.9 \\

7 & $0.39^{+0.02}_{-0.03}$ & $1.57^{+0.01}_{-0.02}$ & $34.93^{+11.23}_{-5.59}$ & $0.64^{+0.08}_{-0.06}$ & $< 1.15$ & $1.34^{+0.11}_{-0.07}$ & $0.07^{+0.05}_{-0.02}$ & $0.46^{+0.05}_{-0.05}$ & $> 0.38$ & 0.4 \\

8 & $0.39^{+0.04}_{-0.04}$ & $1.49^{+0.06}_{-0.07}$ & $28.41^{+15.94}_{-6.46}$ & $0.42^{+0.17}_{-0.13}$ & $< 1.49$ & $1.25^{+0.32}_{-0.08}$ & $0.09^{+0.17}_{-0.02}$ & $0.53^{+0.07}_{-0.12}$ & $> 0.19$ & 0.6 \\

9 & $0.41^{+0.02}_{-0.03}$ & $1.59^{+0.02}_{-0.01}$ & $34.11^{+12.09}_{-5.75}$ & $0.63^{+0.08}_{-0.06}$ & $< 1.19$ & $1.24^{+0.11}_{-0.06}$ & $0.07^{+0.06}_{-0.01}$ & $0.45^{+0.07}_{-0.06}$ & $> 0.35$ & 0.7 \\

10 & $0.39^{+0.04}_{-0.04}$ & $1.61^{+0.02}_{-0.01}$ & $48.71^{+32.88}_{-14.09}$ & $> 0.81$ & $< 1.35$ & $1.24^{+0.17}_{-0.08}$ & $0.21^{+0.02}_{-0.02}$ & $0.63^{+0.17}_{-0.18}$ & $> 0.07$ & 0.8\\

11 & $0.42^{+0.03}_{-0.04}$ & $1.59^{+0.02}_{-0.01}$ & $33.37^{+9.94}_{-5.20}$ & $0.77^{+0.09}_{-0.07}$ & $< 1.19$ & $1.25^{+0.12}_{-0.08}$ & $0.11^{+0.09}_{-0.02}$ & $0.45^{+0.08}_{-0.09}$ & $> 0.26$ & 0.6 \\

12 & $0.38^{+0.05}_{-0.04}$ & $1.59^{+0.01}_{-0.02}$ & $43.46^{+20.51}_{-6.82}$ & $> 0.81$ & $< 1.97$ & $1.002^{+0.12}_{-0.06}$ & $0.02^{+0.25}_{-0.01}$ & $0.51^{+0.07}_{-0.06}$ & $>0.15$ & 0.7 \\

13 & $0.41^{+0.03}_{-0.02}$ & $1.72^{+0.01}_{-0.01}$ & $> 74.27$ & $0.38^{+0.03}_{-0.13}$ & $< 1.44$ & $1.002^{+0.249}_{-0.107}$ & $0.02^{+0.03}_{-0.02}$ & $0.38^{+0.02}_{-0.03}$ & $2.15^{+0.62}_{-0.53}$ & 0.7 \\

14 & $0.47^{+0.03}_{-0.03}$ & $1.77^{+0.03}_{-0.02}$ & $40.89^{+25.93}_{-9.37}$ & $0.21^{+0.04}_{-0.02}$ & $< 1.43$ & $0.93^{+0.14}_{-0.08}$ & $0.02^{+0.04}_{-0.02}$ & $0.39^{+0.02}_{-0.03}$ & $>3.63$ & 0.6 \\

15 & $0.41^{+0.03}_{-0.03}$ & $1.74^{+0.01}_{-0.02}$ & $> 51.37$ & $0.38^{+0.05}_{-0.09}$ & $< 1.56 $ & $0.95^{+0.16}_{-0.06}$ & $0.02^{+0.05}_{-0.03}$ & $0.36^{+0.02}_{-0.03}$ & $2.95^{+1.16}_{-0.72}$ & 0.5 \\

16 & $0.52^{+0.03}_{-0.03}$ & $1.81^{+0.03}_{-0.02}$ & $> 44.91$ & $0.20^{+0.05}_{-0.04}$ & $< 1.87$ & $0.84^{+0.23}_{-0.06}$ & $0.01^{+0.10}_{-0.02}$ & $0.37^{+0.02}_{-0.03}$ & $5.38^{+2.04}_{-2.50}$ & 0.5 \\

17 & $0.48^{+0.03}_{-0.03}$ & $1.77^{+0.03}_{-0.02}$ & $50.42^{+55.34}_{-14.69}$ & $0.23^{+0.06}_{-0.04}$ & $< 1.82$ & $0.86^{+0.21}_{-0.05}$ & $0.02^{+0.09}_{-0.02}$ & $0.35^{+0.02}_{-0.03}$ & $5.70^{+3.12}_{-1.41}$ & 0.5 \\

18 & $0.41^{+0.03}_{-0.03}$ & $1.70^{+0.09}_{-0.05}$ & $34.09^{+11.52}_{-5.92}$ & $0.22^{+0.15}_{-0.14}$ & $1.37^{+0.18}_{-0.23}$ & $1.39^{+0.13}_{-0.19}$ & $0.11^{+0.06}_{-0.03}$ & $0.38^{+0.03}_{-0.03}$ & $2.46^{+0.96}_{-0.70}$ & 0.5 \\

19 & $0.38^{+0.03}_{-0.03}$ & $1.68^{+0.02}_{-0.02}$ & $46.58^{+46.81}_{-12.59}$ & $0.27^{+0.08}_{-0.09}$ & $1.53^{+0.21}_{-0.29}$ & $1.46^{+0.31}_{-0.33}$ & $0.12^{+0.07}_{-0.08}$ & $0.36^{+0.04}_{-0.03}$ & $2.58^{+1.17}_{-0.65}$ & 0.5 \\

20 & $0.34^{+0.03}_{-0.02}$ & $1.66^{+0.02}_{-0.02}$ & $> 29.41^{+6.41}_{-3.93}$ & $0.39^{+0.05}_{-0.04}$ & $< 1.12$ & $1.27^{+0.07}_{-0.05}$ & $0.03^{+0.01}_{-0.01}$ & $0.39^{+0.03}_{-0.02}$ & $1.21^{+0.45}_{-0.32}$ & 0.5 \\

21 & $0.59^{+0.04}_{-0.03}$ & $1.89^{+0.03}_{-0.03}$ & $38.52^{+21.45}_{-8.82}$ & $0.15^{+0.02}_{-0.02}$ & $< 2.24$ & $0.81^{+0.34}_{-0.05}$ & $0.02^{+0.24}_{-0.03}$ & $0.41^{+0.02}_{-0.05}$ & $5.55^{+4.13}_{-1.28}$ & 0.5 \\

22 & $0.63^{+0.03}_{-0.02}$ & $2.01^{+0.03}_{-0.04}$ & $>48.99$  & $0.09^{+0.01}_{-0.02}$ & $< 2.79$ & $0.74^{+0.39}_{-0.05}$ & $0.02^{+0.32}_{-0.01}$ & $0.43^{+0.02}_{-0.03}$ & $6.07^{+1.91}_{-1.36}$ & 0.5 \\

23 & $0.56^{+0.03}_{-0.02}$ & $1.90^{+0.05}_{-0.05}$ & $16.43^{+2.12}_{-1.79}$ & $0.08^{+0.01}_{-0.01}$ & $1.83^{+1.26}_{-0.45}$ & $1.01^{+0.47}_{-0.23}$ & $> 0.12$ & $0.48^{+0.05}_{-0.11}$ & $3.03^{+2.29}_{-0.97}$ & 0.9 \\

24 & $0.49^{+0.03}_{-0.03}$ & $2.00^{+0.05}_{-0.04}$ & $21.55^{+5.29}_{-2.92}$ & $0.15^{+0.03}_{-0.02}$ & $< 1.82$ & $0.97^{+0.14}_{-0.10}$ & $0.04^{+0.02}_{-0.03}$ & $0.53^{+0.02}_{-0.04}$ & $1.64^{+0.48}_{-0.31}$ & 1.1 \\

25 & $0.61^{+0.03}_{-0.02}$ & $2.08^{+0.07}_{-0.08}$ & $> 27.19$ & $0.12^{+0.05}_{-0.02}$ & $2.91^{+0.37}_{-1.28}$ & $1.18^{+0.45}_{-0.35}$ & $> 0.37$ & $0.41^{+0.03}_{-0.08}$ & $> 5.63$ & 0.6 \\

26 & $0.64^{+0.03}_{-0.03}$ & $2.10^{+0.04}_{-0.07}$ & $> 32.37$ & $0.09^{+0.03}_{-0.02}$ & $3.11^{+0.10}_{-0.41}$ & $1.29^{+0.32}_{-0.31}$ & $> 0.72$ & $0.34^{+0.05}_{-0.09}$ & $> 11.49$ & 0.7 \\

27 & $0.57^{+0.03}_{-0.02}$ & $2.04^{+0.07}_{-0.08}$ & $12.41^{+1.40}_{-1.13}$ & $0.12^{+0.03}_{-0.02}$ & $2.16^{+0.41}_{-0.83}$ & $1.31^{+0.29}_{-0.29}$ & $0.42^{+0.13}_{-0.09}$ & $0.43^{+0.07}_{-0.12}$ & $> 6.59$ & 0.6 \\

28 & $0.53^{+0.03}_{-0.03}$ & $2.16^{+0.11}_{-0.10}$ & $16.24^{+5.61}_{-3.082}$ & $0.07^{+0.02}_{-0.02}$ & $1.64^{+1.08}_{-0.34}$ & $1.07^{+0.13}_{-0.18}$ & $0.12^{+0.33}_{-0.09}$ & $0.49^{+0.04}_{-0.02}$ & $2.67^{+1.45}_{-0.45}$ & 0.8 \\

29 & $0.61^{+0.03}_{-0.03}$ & $2.29^{+0.04}_{-0.15}$ & $> 71.18$ & $0.07^{+0.01}_{-0.03}$ & $3.37^{+0.47}_{-1.17}$ & $1.23^{+0.54}_{-0.41}$ & $> 0.42$ & $0.42^{+0.05}_{-0.05}$ & $5.71^{+4.89}_{-2.21}$ & 0.8 \\

30 & $0.57^{+0.03}_{-0.03}$ & $2.28^{+0.06}_{-0.12}$ & $> 27.81$ & $0.12^{+0.05}_{-0.04}$ & $2.64^{+0.71}_{-1.24}$ & $1.15^{+0.29}_{-0.21}$ & $> 0.17$ & $0.48^{+0.06}_{-0.06}$ & $3.15^{+2.64}_{-1.01}$ & 0.8 \\

31 & $0.60^{+0.03}_{-0.02}$ & $2.23^{+0.03}_{-0.04}$ & $> 74.03$ & $0.09^{+0.01}_{-0.03}$ & $3.06^{+0.20}_{-0.67}$ & $1.41^{+0.34}_{-0.27}$ & $> 0.24$ & $0.42^{+0.06}_{-0.06}$ & $> 3.81$ & 0.8 \\

32 & $0.62^{+0.03}_{-0.03}$ & $2.25^{+0.02}_{-0.08}$ & $> 53.71$ & $0.12^{+0.01}_{-0.04}$ & $3.10^{+0.23}_{-0.45}$ & $1.19^{+0.27}_{-0.24}$ & $> 0.74$ & $0.37^{+0.08}_{-0.11}$ & $> 3.95$ & 0.6 \\

33 & $0.59^{+0.03}_{-0.03}$ & $2.18^{+0.08}_{-0.09}$ & $21.42^{+9.11}_{-4.82}$ & $0.09^{+0.03}_{-0.02}$ & $2.29^{+0.67}_{-0.91}$ & $1.03^{+0.38}_{-0.31}$ & $0.44^{+0.04}_{-0.07}$ & $0.42^{+0.05}_{-0.06}$ & $> 5.36$ & 0.6 \\

34 & $0.65^{+0.03}_{-0.03}$ & $1.74^{+0.09}_{-0.07}$ & $10.97^{+1.28}_{-0.73}$ & $0.02^{+0.01}_{-0.01}$ & $3.84^{+0.16}_{-0.74}$ & $> 1.45$ & $> 0.46$ & $0.41^{+0.05}_{-0.02}$ & $7.17^{+2.72}_{-2.80}$ & 0.6\\

35 & $0.50^{+0.02}_{-0.03}$ & $1.80^{+0.05}_{-0.06}$ & $9.78^{+0.59}_{-0.53}$ & $0.06^{+0.01}_{-0.01}$ & $< 1.55$ & $1.14^{+0.11}_{-0.05}$ & $0.03^{+0.04}_{-0.01}$ & $0.57^{+0.02}_{-0.04}$ & $1.68^{+0.49}_{-0.21}$ & 1.06 \\

36 & $0.61^{+0.02}_{-0.03}$ & $2.21^{+0.04}_{-0.06}$ & $> 48.09$ & $0.10^{+0.01}_{-0.03}$ & $2.43^{+0.65}_{-0.96}$ & $1.04^{+0.31}_{-0.26}$ & $0.33^{+0.08}_{-0.11}$ & $0.43^{+0.04}_{-0.05}$ & $5.94^{+2.58}_{-2.90}$ & 0.6 \\

37 & $0.58^{+0.03}_{-0.03}$ & $2.17^{+0.03}_{-0.05}$ & $> 51.26$ & $0.12^{+0.01}_{-0.03}$ & $1.21^{+1.04}_{-0.50}$ & $0.92^{+0.23}_{-0.12}$ & $0.07^{+0.29}_{-0.06}$ & $0.41^{+0.01}_{-0.04}$ & $2.71^{+0.26}_{-0.45}$ & 0.8 \\

38 & $0.58^{+0.03}_{-0.02}$ & $2.23^{+0.03}_{-0.08}$ & $> 38.45$ & $0.12^{+0.09}_{-0.04}$ & $> 2.44$ & $0.92^{+0.23}_{-0.19}$ & $0.10^{+0.30}_{-0.08}$ & $0.45^{+0.03}_{-0.04}$ & $4.41^{+1.45}_{-0.98}$ & 0.6 \\

39 & $0.53^{+0.02}_{-0.01}$ & $2.12^{+0.07}_{-0.06}$ & $19.64^{+6.27}_{-3.49}$ & $0.14^{+0.02}_{-0.01}$ & $< 1.72$ & $1.002^{+0.101}_{-0.067}$ & $0.02^{+0.13}_{-0.02}$ & $0.52^{+0.01}_{-0.03}$ & $2.49^{+0.59}_{-0.31}$ & 0.8 \\

40 & $0.55^{+0.03}_{-0.03}$ & $2.15^{+0.08}_{-0.08}$ & $19.24^{+7.23}_{-3.85}$ & $0.08^{+0.02}_{-0.01}$ & $< 2.15$ & $0.99^{+0.21}_{-0.12}$ & $0.04^{+0.24}_{-0.02}$ & $0.51^{+0.03}_{-0.04}$ & $2.86^{+0.71}_{-0.59}$ & 0.7 \\

41 & $0.71^{+0.02}_{-0.02}$ & $1.91^{+0.07}_{-0.08}$ & $12.84^{+1.86}_{-1.37}$ & $0.05^{+0.01}_{-0.01}$ & $> 1.83$ & $1.31^{+0.35}_{-0.31}$ & $0.40^{+0.23}_{-0.22}$ & $0.42^{+0.05}_{-0.05}$ & $3.26^{+1.10}_{-0.29}$ & 1.2 \\

42 & $0.55^{+0.02}_{-0.03}$ & $1.77^{+0.03}_{-0.02}$ & $28.92^{+6.20}_{-3.79}$ & $0.21^{+0.03}_{-0.02}$ & $< 1.63$ & $0.93^{+0.17}_{-0.03}$ & $0.03^{+0.12}_{-0.02}$ & $0.43^{+0.02}_{-0.03}$ & $< 6.23$ & 0.5 \\

43 & $0.52^{+0.02}_{-0.04}$ & $1.74^{+0.03}_{-0.02}$ & $32.31^{+8.97}_{-5.07}$ & $0.22^{+0.03}_{-0.02}$ & $< 1.69$ & $0.90^{+0.28}_{-0.06}$ & $0.02^{+0.13}_{-0.02}$ & $0.39^{+0.02}_{-0.05}$ & $< 5.47$ & 0.4 \\

44 & $0.49^{+0.03}_{-0.03}$ & $1.71^{+0.02}_{-0.03}$ & $26.83^{+4.76}_{-3.28}$ & $0.18^{+0.03}_{-0.02}$ & $< 1.47$ & $0.96^{+0.07}_{-0.06}$ & $0.02^{+0.02}_{-0.03}$ & $0.39^{+0.02}_{-0.01}$ & $4.15^{+1.66}_{-0.69}$ & 0.5 \\

45 & $0.51^{+0.03}_{-0.03}$ & $1.73^{+0.03}_{-0.01}$ & $24.20^{+3.01}_{-2.45}$ & $0.17^{+0.02}_{-0.02}$ & $<1.41$ & $0.97^{+0.14}_{-0.06}$ & $0.02^{+0.05}_{-0.02}$ & $0.41^{+0.02}_{-0.01}$ & $4.02^{+1.35}_{-0.81}$ & 0.5 \\

46 & $0.54^{+0.03}_{-0.04}$ & $1.79^{+0.03}_{-0.03}$ & $29.76^{+7.28}_{-4.59}$ & $0.16^{+0.02}_{-0.01}$ & $> 1.65$ & $0.91^{+0.09}_{-0.07}$ & $0.03^{+0.18}_{-0.02}$ & $0.45^{+0.02}_{-0.03}$ & $3.74^{+1.68}_{-2.12}$ & 0.6 \\

47 & $0.48^{+0.03}_{-0.03}$ & $1.72^{+0.02}_{-0.03}$ & $28.56^{+6.08}_{-3.84}$ & $0.17^{+0.02}_{-0.01}$ & $> 1.51$ & $0.95^{+0.16}_{-0.06}$ & $0.02^{+0.06}_{-0.02}$ & $0.39^{+0.01}_{-0.02}$ & $3.99^{+1.44}_{-0.82}$ & 0.5 \\

48 & $0.49^{+0.03}_{-0.03}$ & $1.74^{+0.02}_{-0.03}$ & $29.90^{+7.02}_{-4.30}$ & $0.20^{+0.02}_{-0.02}$ & $>1.51$ & $0.94^{+0.17}_{-0.07}$ & $0.02^{+0.08}_{-0.02}$ & $0.41^{+0.02}_{-0.03}$ & $4.14^{+0.05}_{-2.00}$ & 0.4 \\

49 & $0.51^{+0.01}_{-0.02}$ & $1.72^{+0.02}_{-0.02}$ & $32.89^{+10.13}_{-5.25}$ & $0.19^{+0.03}_{-0.02}$ & $< 1.59$ & $0.91^{+0.21}_{-0.06}$ & $0.02^{+0.04}_{-0.02}$ & $0.38^{+0.02}_{-0.03}$ & $5.78^{+1.95}_{-2.18}$ & 0.6 \\

50 & $0.48^{+0.03}_{-0.03}$ & $1.70^{+0.01}_{-0.02}$ & $31.82^{+7.91}_{-5.49}$ & $0.21^{+0.02}_{-0.03}$ & $< 1.53$ & $0.92^{+0.02}_{-0.06}$ & $0.02^{+0.07}_{-0.02}$ & $0.38^{+0.02}_{-0.04}$ & $>4.46$ & 0.4 \\

\bottomrule  
\end{tabular}  
\label{table:tableA2}
\end{minipage}  
\end{table*}

\bibliographystyle{elsarticle-harv} 
\bibliography{bibliography}






\end{document}